\documentclass[tightenlines,aps,twocolumn,showpacs,floatfix]{revtex4}
\usepackage{psfig,float}

\newcommand{\ba}{\begin{eqnarray}}
\newcommand{\ea}{\end{eqnarray}}

\newcommand{\be}{\begin{equation}}
\newcommand{\ee}{\end{equation}}

\begin{document}

\title{%
\hfill{\normalsize\vbox{%
\hbox{\rm UH-511-1029-03}
\hbox{\rm SLAC-PUB-10000} 
}}
\vspace*{0.5cm}

Dilepton rapidity distribution in the 
Drell-Yan process at NNLO in QCD} 

\author{Charalampos Anastasiou,${\!\!}^1$ Lance Dixon,${\!}^1$
Kirill Melnikov,${\!}^2$ and Frank Petriello$^1$}
\affiliation{  
$^1$Stanford Linear Accelerator Center, 
Stanford University, Stanford, CA 94309, U.S.A. \\
$^2$Department of Physics \& Astronomy      
University of Hawaii,      
Honolulu, HI 96822, U.S.A.}    

\begin{abstract}
We compute the rapidity distribution of the virtual photon produced in the
Drell-Yan process through next-to-next-to-leading order in perturbative
QCD.  We introduce a powerful new method for calculating differential
distributions in hard scattering processes.  This method is based upon a
generalization of the optical theorem; it allows the integration-by-parts
technology developed for multi-loop diagrams to be applied to 
{\it non-inclusive} phase-space integrals, and permits a high degree of
automation.  We apply our results to the analysis of fixed target 
experiments.
\end{abstract}


\maketitle

The production of lepton pairs in hadronic collisions, known as the
Drell-Yan (DY) process~\cite{drellyan}, was the first application of
parton model ideas beyond deep inelastic scattering.  Due to its clean
theoretical interpretation in terms of quark-antiquark annihilation into
a vector boson, and large event rates, the DY process has been studied
extensively, and will continue to be investigated at both the Tevatron and
the LHC. The DY process provides valuable information about parton
distribution functions (pdfs), enables measurements of the production rates 
and masses of $W$ and $Z$ bosons, and furnishes a sensitive test for many
varieties of new physics, such as the additional gauge bosons that appear
in many extensions of the Standard Model.  It will also be used for
the more prosaic purpose of monitoring partonic luminosities at the LHC.

Despite the importance of the DY process and the significant amount 
of work devoted to its description, the calculation of higher order 
QCD corrections has proceeded slowly. The next-to-leading order (NLO) QCD
corrections to the total cross section, and the $x_F$ and rapidity
distributions, were calculated nearly 25 years ago~\cite{alt}; the
NNLO corrections to the total cross section were obtained eleven years
later~\cite{Hamberg:1990np}.  No complete calculation of the NNLO QCD
corrections to any differential distribution has been performed,
although partial results exist~\cite{Rijken:1994sh}.

Recently, the NNLO virtual corrections to several interesting hard
scattering processes in QCD have been computed~\cite{glover}; however, the
calculation of real emission contributions, required for complete NNLO
predictions, is still in progress.  These
contributions entail a careful analysis of perturbative multiparticle final
states in generic hard scattering events.  While it is certainly useful to
solve this problem in complete generality, it is also useful to study
specific examples, especially those most urgently needed in experimental
analyses.  It is possible to develop alternative methods of calculation
which can be used to compute basic differential distributions.  In
Refs.~\cite{Anastasiou:2002yz,Anastasiou:2002qz} it was shown how to combine the optical 
theorem with multi-loop computational methods to compute phase space integrals.  
In this Letter we present a non-trivial
application of these ideas; we compute the rapidity distribution of the
virtual photon produced in the DY process through NNLO in perturbative
QCD.

We shall apply our results to the production of lepton pairs in
proton-proton ($pp$) collisions at center-of-mass energies accessible to
fixed target experiments.  The most recent measurements come from Fermilab
experiment E866/NuSea, which measured the dimuon production cross section
in $pp$ and proton-deuteron collisions at $\sqrt{s} \approx 40~{\rm GeV}$
for muon invariant masses in the interval 
$4-16~{\rm GeV}$~\cite{Webb:2003ps}. 
These experiments are sensitive to both the $x \to 1$ components of the
valence quark distribution functions and the moderate $x$ components of
the sea quark distribution functions of the proton.  Neither of these
kinematic configurations are well constrained by other data, so the E866
measurements provide valuable input to a global pdf fit.  The precision of
the E866 measurement is better than 10\% per bin.  Given the significant
($\sim 40$\%) NLO corrections at such energies, the complete NNLO
computation is required.  Although in principle both photon and $Z$ boson
exchanges contribute to this process, the $Z$ exchange component is
suppressed by $M^2 /M_{Z}^2$, where $M$ is the invariant mass of the
lepton pair.  This effect is approximately 1\% for the relevant invariant
masses, and will be neglected in our analysis.

The NNLO calculation is quite challenging technically.  Existing
techniques for computing phase-space integrals are incapable of
handling problems of this complexity.  We introduce here a powerful new
method:  we extend the optical theorem in such a way that the calculation
of differential distributions becomes possible using techniques
developed for multi-loop calculations. To achieve this, we represent the
rapidity constraint by an effective ``propagator.''  This propagator is
constructed so that when the imaginary part of the forward scattering
amplitude is computed using the optical theorem, the ``mass-shell'' 
constraint for the ``particle'' described by this propagator is equivalent 
to the rapidity constraint in the phase space integration.  
We then use the methods described in Ref.~\cite{Anastasiou:2002yz} for 
computing inclusive cross sections, keeping the fake particle propagator 
in the loop integrals, and deriving the rapidity distribution as the 
imaginary part of the forward scattering amplitude.

The production of a lepton pair in a high-energy hadronic collision occurs 
in two distinct steps: first the quarks and gluons from the colliding
hadrons annihilate to create a highly virtual time-like photon;
then the photon decays into a pair of leptons.  In the center-of-mass frame,
the two colliding hadrons have momenta 
$P_{1,2} = \sqrt{s}/2 \left(1,{\bf 0}_{\perp},\pm 1 \right )$.  
A virtual photon of invariant mass $M$ produced in the collision has 
momentum 
$P_\gamma = \left (E,{\bf p}_\perp,p_z \right )$.  
Its energy and momentum are related by the ``mass-shell'' condition 
$E^2 - {\bf p}_\perp^2 -p_z^2 = M^2$, while its
rapidity is defined as 
$\displaystyle
Y = \frac{1}{2} \ln \left (\frac{E+p_z}{E-p_z} \right ). 
$

We first compute the partonic hard scattering cross sections, and then
convolute them with the pdfs of the colliding hadrons.  
The partonic rapidity distributions for the hard scattering of
partons $i,j$, with momentum $p_1 = x_1P_1$ and $p_2 = x_2 P_2$
respectively, are obtained by integrating the hard scattering matrix
elements over the phase-space of the final state particles with the
rapidity and mass of the virtual photon kept fixed: \be \frac{{\rm
d}\sigma_{ij}}{2e^{2Y}{\rm dY}} =\int {\rm d}\Pi_f |{\cal M}_{ij}|^2
\delta \left (e^{2Y} - \frac{E+p_z}{E-p_z} \right ).  \ee
The rapidity constraint can be rewritten using the incoming parton
momenta as
\be
\delta \left (e^{2Y} - \frac{E+p_z}{E-p_z}
\right ) = e^{-2Y} \delta \left( 
    \frac{P_\gamma \cdot [p_1 - u p_2] }{P_\gamma \cdot p_1}
                          \right),
\label{eq4}
\ee
where $u= \frac{x_1}{x_2}e^{-2Y}$.  
The simple Lorentz boost properties of the rapidity are helpful at
NNLO.  In comparison to the total cross section computation, only 
one additional dimensionless variable, $u$, is introduced.  
Computation of the distribution in $x_F = 2 p_z/\sqrt{s}$, for example, 
would require two new variables.

At leading order in $\alpha_s$, the production of the virtual photon
occurs through the annihilation of a $q\bar{q}$ pair.  Only the
virtual photon is produced in the collision, rendering the phase-space
integrations trivial.  At higher orders in $\alpha_s$, inelastic channels
contribute.  At ${\cal O}(\alpha_s)$, for example, we must consider
also $q \bar q \to \gamma^* g$ and $q g \to q \gamma^*$.  It is still
quite simple to perform these phase space integrations using standard techniques.
At higher orders, however, this approach becomes impractical, so we 
adopt instead the method of Ref.~\cite{Anastasiou:2002qz}, which
can be applied efficiently at NNLO.

We first represent the $\delta$-function in Eq.~(\ref{eq4}) as the
imaginary part of an effective propagator: 
\be 
\delta(x) \to \frac{1}{2\pi i} 
\left [ \frac{1}{x-i0} - \frac{1}{x+i0} \right ].
\label{eq5}
\ee
Next we map the constrained phase-space integrals onto forward 
scattering loop integrals~\cite{Anastasiou:2002yz}.
We denote the difference of propagators with opposite $i0$ prescription,
such as that shown in Eq.~(\ref{eq5}), by a cut propagator; 
final state particles on mass-shell are also represented by 
cut propagators.  The $\delta$-function constraint~(\ref{eq4}) becomes
an unconventional propagator, linear in the loop momentum.

At NLO, we must consider integrals of the following general form:
\be
I(\nu_1,\nu_2,\ldots,\nu_5) = 
\int \frac{{\rm d}^dk}{(2\pi)^d} \frac{1}{A^{\nu_1}_1 \ldots A^{\nu_5}_5},
\label{eq6}
\ee
where $A_1 = k^2-M^2 \pm i0,~~~A_2 = (k+p_1)^2,~~~A_3 = (k+p_1+p_2)^2 \pm
i0, A_4 = (k+p_2)^2$, and $A_5 = k\cdot p_1 - uk \cdot p_2 \pm i0.$ The
propagators $A_1,A_3$ and $A_5$ should be ``cut'' according to
Eq.~(\ref{eq5}), indicating that the corresponding particles are on-shell.
The propagators $A_{1..5}$ are linearly dependent; we can therefore
eliminate both $A_2$ and $A_4$ from the integrand in Eq.~(\ref{eq6}) by
partial fractioning.  Partial fractioning produces integrals with
either $\nu_1,\nu_3$ or $\nu_5$ equal to zero.  When the cutting 
rule~(\ref{eq5}) is applied, such integrals vanish.  All phase-space 
integrals of the form of Eq.~(\ref{eq6}) can be reduced to a single
``master'' integral, $I(1,0,1,0,1)$. 
The fact that only partial fractioning relations are required to perform
this reduction is specific to NLO; we will discuss a more general 
reduction technique when we consider the NNLO corrections.

We compute the virtual corrections to the leading order production process
$q \bar q \to \gamma^*$ in the standard fashion, since the rapidity
constraint leaves this calculation unaffected.  After combining the real
and virtual corrections and performing the collinear factorization, we
arrive at the LO and NLO results for the partonic rapidity 
distributions~\cite{alt}, which we present here for completeness.

We write the partonic differential cross section for the process $
i+j \to \gamma^* X$, renormalized in the ${\overline {\rm MS}}$ scheme, 
as
\be
(1-z)\frac{{\rm d} \sigma_{ij}}{{\rm dY}} =  \eta^{(0)}_{ij} 
+ \left ( \frac{\alpha_s}{\pi} \right ) \eta^{(1)}_{ij}
+ \left ( \frac{\alpha_s}{\pi} \right )^2 \eta^{(2)}_{ij} 
+{\cal O}(\alpha_s^3),
\ee
where $\alpha_s = \alpha_s(M)$ is the strong coupling 
constant assuming $n_f$ massless quark flavors, renormalized 
at the scale $M$. The factorization scale is also set equal to $M$;
the dependence on both the renormalization and the factorization scales can be 
restored by using the renormalization group invariance of the 
hadronic cross section.

At the lowest order in $\alpha_s$, the virtual photon 
can be produced only in the collision of a quark and antiquark of the same flavor. Therefore,
\be
\eta^{(0)}_{ij} = Q_q^2 \left ( \delta_{iq}\delta_{\bar q j} 
+ \delta_{i \bar q} \delta_{q j} \right )
\delta(1-z) \left ( \delta(y) + \delta(1-y) \right),
\ee
where $z = M^2/\hat{s}$, $\hat{s}$ is the partonic 
Mandelstam invariant and $y = (u-z)/(1-z)/(1+u)$.

At NLO, the $q \bar q$ channel receives ${\cal O}(\alpha_s)$ corrections, 
and the $qg$ and $\bar{q}g$ channels contribute.  We find
\ba
&&\frac{\eta^{(1)}_{q \bar q}}{Q_q^2} = 
{8 \over 3} { z^2 \over 1+z } \biggl\{
[\delta(y)+\delta(1-y)] \biggl[
\delta(1-z) ( 2\zeta_2 - 4 )
\nonumber \\
&& 
+ 4 \biggl[ { \ln(1-z) \over 1-z } \biggr]_+
- 2 (1+z) \ln(1-z)
- { 1+z^2 \over 1-z } \ln z
\nonumber \\
&& 
 + 1 - z \biggr]
+ \biggl( 1 + {(1-z)^2 \over z} y(1-y) \biggr)
\nonumber \\
&& 
\times
\biggl[ { 1+z^2 \over [1-z]_+ } 
   \biggl( {1\over y_+} + { 1 \over [1-y]_+ } \biggr)
  - 2 (1-z) \biggr] \biggl\} 
,
\label{qqNLO}
\ea
%
%
\ba
&&\frac{\eta^{(1)}_{q g}}{Q_q^2} =
{ z^2 \over 1+z } \biggl\{
\delta(y) \biggl[ 
[ z^2 + (1-z)^2 ] \ln \frac{(1-z)^2}{z} 
\nonumber \\
&& 
+ 2 z (1-z)
\biggr]
+ \biggl( 1 + {(1-z)^2 \over z} y(1-y) \biggr)
\nonumber \\
&& 
\times \biggl[ 
[ z^2 + (1-z)^2 ] {1\over y_+}
+ 2 z (1-z) + (1-z)^2 y \biggr] \biggl\}
\label{qgNLO}.
\ea
%
%
Results for the other channels are obtained by permuting 
partonic labels and changing $y \to 1-y$ in 
Eqs.~(\ref{qqNLO}, \ref{qgNLO}).

We now discuss the calculation of the NNLO contributions.  The purely
virtual correction to the rapidity distribution is the same as the virtual
correction to the total cross section, and is straightforward to compute
using standard techniques.  We compute both the real-virtual and the
real-real corrections using the method described above.  However, to
achieve a complete reduction to master integrals at NNLO we must
supplement the partial fractioning identities with integration-by-parts
(IBP) identities~\cite{ibps} typically used in the reduction of loop
integrals.  Our substitution of the rapidity constraint with an effective
propagator facilitates the use of IBP identities in phase-space
integrals.  The combined system of equations reduces all required
integrals to approximately thirty master integrals, which depend on two
variables, $u$ and $z$.  The equations are also used to construct 
differential equations satisfied by the master integrals,
following Refs.~\cite{Anastasiou:2002yz,Gehrmann:1999as}.  Here, two
first order inhomogeneous partial differential equations are derived for
each integral.  These equations are solved, and the boundary conditions
obtained by considering simple kinematic limits.

At NNLO, the following partonic channels contribute: $q\bar{q}$, the
scattering of a quark and anti-quark of the same flavor; $q (\bar{q})g$;
$gg$; and $q_i q_j (\bar{q}_j)$, the scattering of quarks (anti-quarks) of
arbitrary flavor.  The complete analytic results for the partonic cross
sections are quite lengthy, and will be presented elsewhere.

Integrating the partonic cross sections over the virtual photon rapidity,
we reproduce the ${\cal O}(\alpha_s^2)$ correction to the total cross
section computed in Ref.~\cite{Hamberg:1990np}.  This provides a strong
check on our result.

Finally, we must convolute the renormalized partonic cross sections with
the partonic structure functions to obtain the experimentally measured
cross section.  We consider the doubly differential cross section 
${\rm d}^2\sigma/{\rm d}M {\rm d}Y$:
$$
\frac {{\rm d}^2 \sigma}{{\rm d} M {\rm d} Y} = 
\frac{4\pi \alpha^2}{9 M^3} 
\sum_{i,j} \int {\rm d}x_1 {\rm d}x_2 f_i(x_1) f_j(x_2)
\frac{{\rm d}\sigma_{ij}(z,u)}{{\rm d}Y},
$$
where $z = M^2/(sx_1x_2)$, $u = (x_1/x_2) e^{-2Y}$, 
and $\alpha$ is the electromagnetic coupling 
evaluated at the scale $M$; numerically, $\alpha^{-1} \approx 132$.  
We use a consistent set of pdfs and $\alpha_s$ at each order;  
the `NNLO' set relies on an approximate set of NNLO 
splitting kernels~\cite{martin}.

%
\noindent
\begin{figure}[htbp]
\vspace{0.1cm}
\centerline{
\psfig{figure=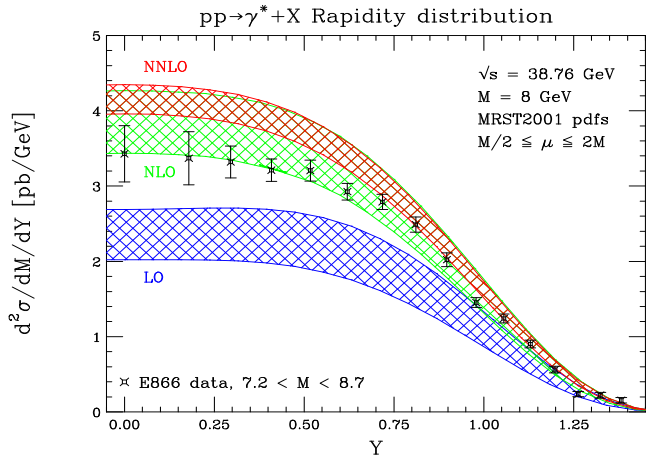,height=6.0cm,width=8.5cm,angle=0}}
\caption{The CMS rapidity distribution of the lepton pair produced 
in $pp$ collisions at LO (lower band), NLO (middle band), and 
NNLO (upper band), for parameter choices relevant for fixed target 
experiments, along with E866 data.  The upper (lower) edge of each band 
denotes the renormalization/factorization scale choice $\mu=M/2$ ($\mu=2M$).} 
\label{rapplot}
\vspace{-0.3cm}
\end{figure}
\vspace{-0.22cm}

In Fig. 1, we present the center-of-mass system (CMS) rapidity
distribution of $M=8$~{\rm GeV} lepton pairs produced in $pp$
collisions at $\sqrt{s} = 38.76$ GeV, along with data from the
fixed-target experiment E866/NuSea~\cite{Webb:2003ps}.
We set the renormalization and factorization scales both equal to $\mu$. 
The LO, NLO, and NNLO results are shown as bands in the figure,
indicating the variation of the cross sections between the 
scale choices $\mu = M/2$
and $\mu = 2M$. 
The NLO and NNLO distributions become more sharply peaked at central 
rapidities; this is due primarily to the evolution of the parton 
distribution functions beyond leading order.  The significant scale 
dependence of the NLO cross section, which reaches nearly 25\% over 
the interval $M/2 \leq \mu \leq 2M$, is reduced to approximately 
10\% at NNLO.  The magnitude of the NNLO corrections depends upon the 
choice of $\mu$; typically, they increase the NLO result by 
approximately 5-15\%.   We note that the NNLO corrections are 
drastically reduced for the scale choice $\mu = M/2$.
The NNLO corrections computed in the so-called ``soft'' approximation, 
which retains only those terms that are singular in the limit $z \to 1$, 
lead to a $\sigma_{\rm NNLO}$ which is lower than the result of the 
full calculation by approximately $20$\%~\cite{Rijken:1994sh}. 

The E866 data points are based on measured $x_F$ 
distributions~\cite{Webb:2003ps}, converted from $x_F$ to $Y$ with
the aid of the $p_\perp$ distribution~\cite{Webb:2003bj}.
A $\pm 6.5\%$ normalization error common to all data points 
is not shown.
The data lie somewhat below the NNLO prediction at smaller $Y$,
and rise to meet it at large $Y$, a trend also visible at other values
of $M$.  Recall that the antiquark distribution is 
{\it derived} partly from the DY process, fit to the NLO $x_F$ distribution.
Fitting to our NNLO rapidity distribution instead may result in a smaller
antiquark distribution function at moderate $x$.  Further comparison with 
$pd$ as well as $pp$ data is in progress.

%
\noindent
\begin{figure}[htbp]
\vspace{0.1cm}
\centerline{
\psfig{figure=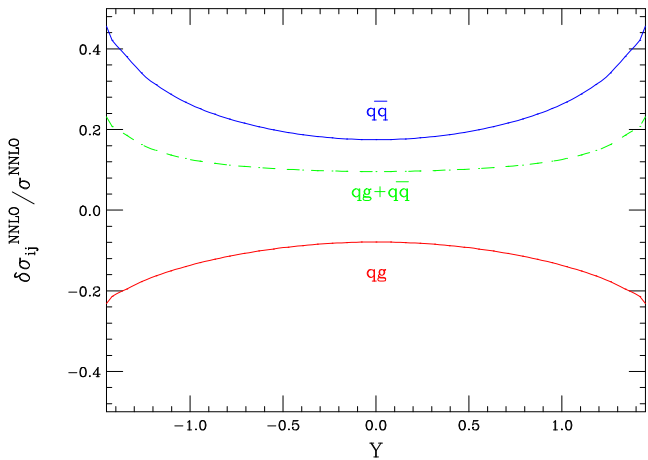,height=6.0cm,width=8.5cm,angle=0}}
\caption{The NNLO corrections for the partonic channels $q\bar{q}$ and 
$qg$, normalized to the complete NNLO differential cross section,  
for $\sqrt{s} = 38.76$ GeV, $M=8$ GeV, and $\mu=M$.} 
\label{ratplot}
\vspace{-0.3cm}
\end{figure}
\vspace{-0.22cm}

We now separate our result into its partonic components.
The $q\bar{q}$ and $qg$ pieces contribute the majority of the
result; the remaining channels are a factor of 50-100 smaller.  The
magnitude of the NNLO result is determined by a significant cancellation
between the $q\bar{q}$ and $qg$ channels.  We illustrate this cancellation
by plotting the NNLO contributions of these channels, together with their
sum, normalized to the complete NNLO differential cross section in Fig. 2.
The sum of these channels also contributes a much flatter correction to
the rapidity distribution than either piece individually.  The $qg$
channel contributes a significant fraction of the complete differential
cross section; combining both the NLO and NNLO $qg$ pieces, we find that
they account for about 15\% of the complete NNLO result for central
rapidities, and nearly 40\% for larger ($Y \ge 1$) rapidities.  This
indicates that the NNLO rapidity distribution is quite sensitive to the
gluon content of the colliding protons.

Finally, we discuss the dependence of the perturbative K-factors upon rapidity.
We define the K-factors 
as follows: 
$K^{\rm (N)NLO}(Y)=\sigma^{\rm (N)NLO}/ \sigma^{\rm LO}$, 
and $K^{(2)}(Y) = \sigma^{\rm NNLO}/ \sigma^{\rm NLO}$.  
We present them in Fig. 3.  The significant variation of both 
$K^{\rm NLO}(Y)$ and $K^{\rm NNLO}(Y)$ with rapidity, a nearly 25\% 
change from $Y=0$ to $Y=1$,
illustrates that the LO result provides a poor approximation to the shape
of the rapidity distribution, as does the LO result weighted by the NNLO
K-factor computed from the inclusive cross section.  However, the relative
flatness of $K^{(2)}$ indicates that the NLO result does accurately
predict the shape of the distribution; the NLO differential cross section
weighted by $\sigma^{\rm NNLO}/ \sigma^{\rm NLO}$, the ratio of NNLO and
NLO inclusive cross sections, is valid at these energies to approximately
3-5\%.  This result appears rather promising, since it suggests a simple
and fairly accurate way of incorporating NNLO corrections into NLO Monte
Carlo event generators by renormalizing with constant $K$-factors.  It
remains to be seen, however, if the same conclusion is valid more generally.

\noindent
\begin{figure}[htbp]
\vspace{0.4cm}
\centerline{
\psfig{figure=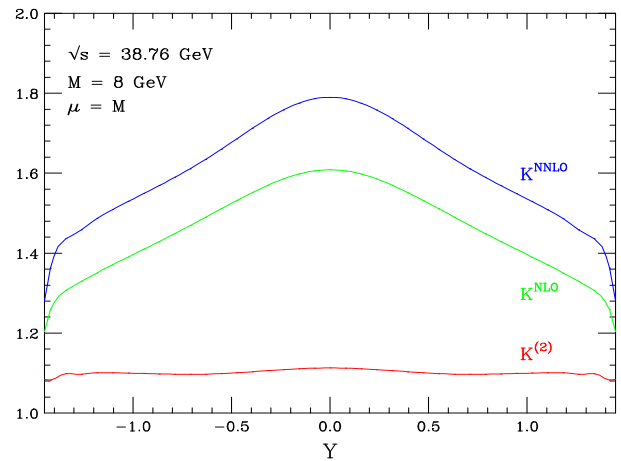,height=6.0cm,width=8.0cm,angle=0}}
\caption{The K-factors $K^{\rm NLO}(Y)=\sigma^{\rm NLO}/ \sigma^{\rm LO}$, 
$K^{\rm NNLO}(Y)=\sigma^{\rm NNLO}/ \sigma^{\rm LO}$, and 
$K^{(2)}(Y) = \sigma^{\rm NNLO}/ \sigma^{\rm NLO}$, for $\mu=M$.} 
\label{scaleplot}
\vspace{-0.3cm}
\end{figure}
\vspace{-0.19cm}

In conclusion, we have described a calculation of the NNLO QCD corrections
to the rapidity distribution in the Drell-Yan process. We have introduced
a powerful new method for the calculation of differential quantities in
perturbation theory.  Although we have presented only a specific example
of this technique, it is clear that this method is of more general
applicability; the relation between differential distributions and forward
scattering amplitudes described above enables the use of multi-loop
technology for the calculation of a large class of phase space integrals.
We are confident that this technique will be succesfully applied to
compute other quantites of phenomenological interest.

{\bf Acknowledgments}: 
This  research is partially supported by the DOE under contracts 
DE-AC03-76SF00515 and DE-FG03-94ER-40833 and by the 
University of Hawaii startup grant.
We thank Mike Leitch and Jason Webb 
for useful discussions.  
C. A. thanks the University of Hawaii for kind hospitality during the 
course of this work.


\end{document}